\documentclass[11pt]{article}

\usepackage{amssymb,latexsym,amsmath}
\usepackage{graphicx}
\usepackage{graphicx,amssymb}
\usepackage{bm}
\usepackage{float}
\usepackage{xcolor}
\usepackage{csquotes}

\begin{document}

\title{Post-Newtonian expansion with Galilean covariance}

\author{G\'ery de Saxc\'e\footnote{Univ. Lille, CNRS, Centrale Lille, UMR 9013 – LaMcube – Laboratoire de m\'ecanique multiphysique multi\'echelle, F-59000, Lille, France}}

\maketitle

\begin{abstract} 
The Galilean gravitation derives from a scalar potential and a vector one. Poisson's equation to determine the scalar potential  has no the expected Galilean covariance. Moreover, there are three missing equations to determine the potential vector. Besides, we require they have the Galilean covariance. These are the issues addressed in the paper. To avoid the drawbacks of the PPN approach and the NCT, we merge them into a new framework. The key idea is to take care that every term of the $c$ expansion of the fields are Galilean covariants or invariants. The expected equations are deduced by variation of the Hilbert-Einstein functional. The contribution of the matter to the functional is derived from Souriau's conformation tensor. We obtain a system of four non linear equations, solved by asymptotic expansion. 
\end{abstract}


\section{Introduction}

General Relativity (GR) is a consistent framework for mechanics and physics of continua. The stress-energy tensor, representing the matter and divergence free, is identified to a tensor linked to the curvature of the space-time manifold. These equations allow to determine the  $10$ components of the metrics which are the potentials for the gravitation. Is this scheme transposable to the classical mechanics? Firstly, we swap Poincar\'e group for Galileo's one. There is no metrics but the Galilean gravitation has $4$ potentials, the classical scalar one $\phi$ and $3$ components of a space vector potential $A$ that generates Coriolis force \cite{AffineMechBook}. 

It is generally admitted that $\phi$ can be determined solving Poisson's equation 
\begin{equation}
\Delta \phi = 4\,\pi\,k_N \rho 
\label{Delta phi = 4 pi k_N rho}
\end{equation}
but it is worth to observe this equation has no the expected Galilean covariance. Moreover, there are $3$ missing equations to determine $A$. Of course, we require they have the Galilean covariance \cite{AffineMechBook}. These are just the issues we would like to address in this paper. In the literature, two theories were proposed to find the equations of the Galilean gravitation field. 

 The Parameterized Post-Newtonian (PPN) theory is the oldest one \cite{Eddington 1922, Will 1971, Ni 1972, Weinberg 1972, Misner 1973, Poisson Will 2014}. The idea is to start with the equations of the General Relativity and to expand the metrics and the derived quantities in terms of a small parameter under the assumption of weak gravitation fields and small velocities. While the gravitation in everyday life is perfectly explained by Newton's theory, they are tiny effects that can be predicted other than by GR. In the approximation of weak fields and small velocities, they can be captured by PPN theory. Although very popular, this approach has the flaw that each terms of the asymptotic expansion of the field equations has not the Galilean covariance.

An alternative approach is Newton-Cartan theory (NCT) \cite{Trautman 1963, Trautman 1966, Dombrowski 1964, Kunzle 1972, Ehlers 1981, Duval 1985, Loos 1985}  that is directly based on the Galilean gravitation and avert the asymptotic expansions but some assumptions are claimed that lead to Poisson's equation then does not have the expected Galilean covariance. Trautman's idea consists in replacing (\ref{Delta phi = 4 pi k_N rho}) by the choice of the following Ricci tensor:
\begin{equation}
   \bm{R}' = 4\,\pi\,k_N \rho\,\bm{\tau} \otimes \bm{\tau}
\label{R' = 4 pi k_N rho tau otimes tau} 
\end{equation}
where $\bm{\tau}$ is the the clock form, one of the elements of the Toupinian structure \cite{Toupin, Noll, Kunzle 1972}. Recently, enhancements of this approach where achieved in different way. In GR, a vielbein is a geometrical object which implements a basis change so as to diagonalize the metric fields. An extended vielbein was proposed for NCT in \cite{Geracie 2015}, replacing Poincar\'e's group by Bargmann's group \cite{Bargmann, Duval Horvathy 2014}. The Galilean gauge theory (GGT) is a gauging procedure of Bargmann algebra in which several curvature constraints are imposed \cite{Andringa Bergshoeff 2011}. Then (\ref{Delta phi = 4 pi k_N rho}) can be derived from (\ref{R' = 4 pi k_N rho tau otimes tau}) by imposing Trautmann and Ehlers conditions. In \cite{Andringa Bergshoeff 2013}, a supersymmetric extension of the three-dimensional Newton-Cartan gravity is constructed by gauging a super-Bargmann algebra. Another approach is the torsional NCT that can be achieved using the GGT as in \cite{Banerjee 2016} or assuming an expansion of the metrics as in the PPN  theory \cite{Bleeken 2017}. Another topic of interest is the coupling between the gravity and the electromagnetism. Two Galilean covariant expressions of Maxwell electrodynamics were derived by Le Bellac and L\'evy-Leblond, respectively in the electric and magnetic limits \cite{Le Bellac Levy-Leblond 1973}. The topic was recently revisited using two distinct $5$-dimensional approaches. In \cite{de Montigny 2003, Santos de Montigny 2004}, Bargmann's group is used while in \cite{Bleeken 2016} it is shown that Kaluza-Klein reduction and the non relativistic limit commute.

To avoid the drawbacks of the PPN approach and the NCT, our approach consists in merging them into a new framework, in the footsteps of pionnering works by Dautcourt \cite{Dautcourt 1997} and Tichy and Flanagan \cite{Tichy Flanagan 2011}. As in \cite{Bleeken 2017}, we use a $c$ expansion of the metric. However, our connection is torsion-free and the point of view is different. Galilean tensors are the extension of the Euclidean tensors to the spacetime. Then we take care to check that every term of the expansion of the familiar tensors of GR are Galilean. This is one of the ways to reconcile the PPN approach and NCT. We do not use explicitly Bargmann's group but it is present in the background, as shown in Section \ref{Section The connection}.

The work is organized as follows. To avoid ambiguity, we present in Section \ref{Section Notations} some notations which are not necessarily usual. In Section \ref{Section Galilean Mechanics}, we recall some basic elements of NCT. In particular, we introduce the Galilean charts and, on the ground of the $G$-structure concept \cite{Kobayashi 1972}, the Galilean charts and the galileomorphisms. Also we derive the potentials of the Galilean gravitation from the closure condition of the presymplectic $2$-form. In Section \ref{Section Galilean tensors}, we introduce the Galilean tensors by restriction of the transformations law of tensors to the linear Galilean transformations. They can be seen as orbits for the action of Galilei group onto the tensor components. In Section \ref{Section Space-time metrics}, we start with a space-time metrics built from the Galilean gravitation as in NCT but without too restrictive assumptions. We verify that each terms of the expansion of the metrics, limited to the first order in $c^{-2}$, are Galilean $2$-covariant tensors. 

Next, we consider the General Relativity as in the PPN theory but in its variational version \cite{Souriau 1964} using Hilbert-Einstein functional and taking care that each term of the asymptotic expansion has the expected Galilean covariance. In Section \ref{Section Functional due to the matter}, we start with the contribution of the matter to the functional.  The motion of a continuum is described by a line bundle of which each fiber represents the trajectory of a material particle. The covariant ansatz is the conformation tensor introduced by Souriau in \cite{Souriau 1964} to generalize in GR the classical concept of deformation.  
Next, we expand Levi-Civita connection (Section \ref{Section The connection}), the Riemann-Christoffel tensor (Section \ref{Section Riemann-Christoffel tensor}), the Ricci tensor (Section \ref{Section Ricci tensor}) and the scalar curvature (Section \ref{Section Scalar curvature}). We verify the scalar curvature is, as expected, a Galilean invariant.  In Section \ref{Section Functional due to the geometry}, the contribution of the geometry to the functional is depending on two unknown parameters that, by identification in a simple situation, are expressed in terms of the the gravitational constant, the speed of the light and the cosmological constant.  By variation of the functional with respect to the potentials of the gravitation, we obtain Einstein equations in the Galilean frame in the form of the equality between two $4$-vectors, one generated by the curvature and the other one from the matter. Considering in Section \ref{Section Asymptotic expansion of the solution} that the motion of the matter is known through the density and velocity fields of the matter, Einstein equations can be used to determine the Galilean gravitation due the presence of the matter. As the equations are non linear, the dominant terms are obtained by using a $c$ expansion of these equations.

\section{Notations}
\label{Section Notations} 

The derivative of a scalar field at $x \in \mathbb{R}^n$, denoted $ D f (x) = \partial f / \partial x$ is a linear map from $\mathbb{R}^n$ into $\mathbb{R}$, that is a $n$-row of which the $i$-th element is the partial derivative of $f$ with respect to $x^i$. Its  gradient is the $n$-column:
$$ grad\,f = \left( \dfrac{\partial f}{\partial x} \right)^T\ .
$$
Let $v$ be a vector field valued in $\mathbb{R}^p$. Its derivative at $x$, denoted $ D v (x) = \partial v / \partial x$ 
is a linear map from $\mathbb{R}^n$ into $\mathbb{R}^p$, that is a $p \times n$ matrix. Its gradient is the $n \times p$ matrix:
$$ grad\,v = \left( \frac{\partial v}{\partial x} \right)^T\ .
$$ 

For a $3$-column $u$, $ j\,(u)$ is the unique skew-symmetric matrix such that $j\,(u)\,v = u \times v$. For any $R \in \mathbb{SO} (3)$, it satisfies the identity: 
\begin{equation}
   j (R^T u) = R^T j(u) R
\label{j(R^T u)} 
\end{equation}
The curl of the $3$-column field $v$ can be defined by:
\begin{equation}
   j (curl\,v) = \frac{\partial v}{\partial x} - \left( \frac{\partial v}{\partial x} \right)^T\ .
\label{defi curl} 
\end{equation}
The divergence of a field $M$ of square matrices of order $n$ is the field $div\,M\in (\mathbb{R}^n)^*$ of $n$-columns such that for every uniform vector field $k (x) = C^{te}\in \mathbb{R}^n$:
$$   (div\,M) \cdot k = div\,(M\,k)\ .
$$

\section{Galilean Mechanics}
\label{Section Galilean Mechanics} 

\subsection{Space-time}
The space-time will be consider as a differential manifold $\mathcal{M}$ of dimension $4$. A point $\bm{X} \in \mathcal{M}$ represents an event. The 4-column vector of its coordinates $(X^{\alpha})_{0 \leq \alpha \leq 3} $ in a choosen local chart will be denoted $X$. The local charts and the associated coordinate systems in which the distances and times are measured will be called Galilean. In such charts, $X^0 = t$ is the time and $X^i = x^i$ for $1 \leq i \leq 3$ are the spatial coordinates, so we can write 
$$ X =\left( \begin{array} {c}
                       t \\
                       x \\
                    \end {array} \right) 
$$
It is worth to remark that we have adopted Souriau's definition in \cite{Souriau 1964} for a chart $\phi:V_\phi \to U_\phi : X \mapsto \bm{X} = \phi (X)$ where $V_\phi$ is an open subset of $\mathbb{R}^n$ and $U_\phi \subset \mathcal{M}$. in the classical definition, a chart is $\phi^{-1}$, which is not a problem since it is an homeomorphism.
                    
\vspace{.5cm} 
             
\subsection{Galilei group and geometry} 
Let $V$ (resp. $V'$) the components of a tangent vector $\overrightarrow{\bm{V}}$ in a local chart chart $X$ (resp. $X'$). The affine transformations $V'  = P V + V_0$, where $P \in \mathbb{GL} (4) $ and $V_0
 \in \mathbb{R}^4$, preserving the distances, the time durations, the uniform straight motions and the oriented volumes are called Galilean transformations. In what follows, we are only interested by the linear ones
\begin{equation}
P =\left( \begin{array} {cc}
                        1  & 0 \\
                        u & R
             \end {array} \right) 
\label{Galilean transfo}
\end{equation}
where $u \in \mathbb{R}^{3}$ is the velocity of transport, or Galilean boost, and $R \in \mathbb{SO} (3)$ is a spatial rotation. The set of all these transformations is a Lie subgroup $\mathbb{GAL}$ of the affine group $\mathbb{GA} \left( 4 \right)$ called Galilei group. It equips the space-time with a structure equivalent to the one proposed by Toupin \cite{Toupin}, taken up later on by Noll \cite{Noll} and K\"unzle \cite{Kunzle 1972}. The toupinian structure of the space-time is based on two canonical tensors, a semi-definite contravariant symmetric tensor $\bm{h}$ of signature $(0+++)$ and a covector $\bm{\tau}$, the clock form such that $h^{\alpha\beta} \tau_\beta =0$. This neoclassic modelling offers a theoretical frame for the universal or absolute time.
\vspace{.5cm}

\subsection{$G$-structure and Galilean charts}

To every variation $dX$ of the coordinates can be associated a tangent vector:
$$ \overrightarrow{d \bm{X}} = \frac{\partial \phi}{\partial X}\ dX = S_\phi  dX 
$$
what defines a basis $(\vec{\bm{e}}_i)$ of $T_{\bm{X}} \mathcal{M}$ as image of the canonical basis of $\mathbb{R}^n$.

Let $\mathcal{M}$ be a differentiable manifold of dimension $n$ and the corresponding principal fibre bundle $\pi: L(\mathcal{M})\rightarrow \mathcal{M}$ of basis with structure group $\mathbb{GL} (n)$. Let $G$ be a Lie subgroup of $\mathbb{GL} (n)$. By a $G$-structure on $\mathcal{M}$, we mean a subbundle $L_G$ of $L (\mathcal{M})$ with structure group $G$ \cite{Kobayashi 1972}. 

A $G$-structure $L_G$ is integrable if every point $\bm{X}$ of $\mathcal{M}$ has a chart $\phi:V_\phi \mapsto U_\phi$ around $\bm{X}$ with local coordinate system $X$ such that the cross-section $\bm{X}\mapsto S_\phi (\bm{X})$ over $U_\phi$ is a cross-section of $L_G$ over $U_\phi$. In other words, $\bm{X}\mapsto S_\phi (\bm{X})$ is a natural frame. Then $X$ is said $G$-admissible. If $X'$ is another $G$-admissible local coordinate system over $U_{\phi'}$, then the Jacobian matrix $\partial X'/\partial X $ belongs to $G$ at each point of $U_\phi \cap U_{\phi'} $:
\begin{equation}
\frac{\partial X'}{\partial X} = P^{-1} \in G\ .
\label{Jacobean matrix} 
\end{equation}
We say that the transition map $X \mapsto X'$ is a $G$-morphism. Although in general $G$-structures are not integrable --in particular in the important case of the Riemannian geometry, the obstruction being the curvature--, it is worth to notice that the Galilean structures are integrable. The Lie group of the linear Galilean transformations (\ref{Galilean transfo}) is denoted $\mathbb{GAL}_0$. The $\mathbb{GAL}_0$-morphisms are called galileomorphisms and are characterized by the following result:

\textbf{Theorem}. Any galileomorphism $X \mapsto X'$ is compound of a rigid body motion and a clock change:
$$
{x}'=(R\,(t))^T\,(x-x_0 \,(t)),
\quad
{t}'=t+ t_0 
$$
where $t\mapsto R\,(t)\in \mathbb{SO}(3)$ and $t\mapsto x_0 (t)\in \mathbb{R}^3$ are smooth mappings, and $t _0 \in \mathbb{R}$ is a constant. Then the velocity of transport is given by:
\begin{equation}
\label{eq29}
u=\omega \,(t)\times \,(x-x_0 \,(t))+\dot {x}_0 \,(t)
\end{equation}
where  $\omega $ is Poisson's vector such that: $\dot {R}=j\,(\omega)\,R$.

\vspace{.5cm}

For the proof, the reader is referred to (\cite{AffineMechBook}, p. 339-341, Theorem 16.4). The idea is to use Frobenius method to integrate the PDE:
$$ \frac{\partial X'}{\partial X} = P^{-1} \in \mathbb{GAL}_0\ .
$$
There exists a family of local charts which are deduced one from each other by such transition maps. We call them Galilean charts or Galilean coordinate systems or Galilean reference frames. In a physical point of view, the importance of these charts lies in the fact that they are the coordinate systems in which the observers measure the durations and distances.

\subsection{Galilean connections and equation of motion}
A covariant differential or connection of a vector field is $ \nabla V^\alpha = d V^\alpha + \Gamma^\alpha_\beta (dX)\ V^\beta$ where, using Christoffel symbols, the elements of the connection matrix $\Gamma$ are $ \Gamma^\alpha_\beta (dX) = \Gamma^\alpha_{\mu\beta} dX^\mu$.  

At each group of transformation is associated a family of connections and the corresponding geometry. We call Galilean connections the symmetric connections on the tangent bundle $T \mathcal{M}$ associated to Galilei group \cite{Toupin, Truesdell, Kunzle 1972}, \textit{i.e.} such that the two canonical tensors of the toupinian structure are parallel-transported. In a Galilean chart, the connection matrix is valued in the Lie algebra $\mathfrak{gal}_0$ of $\mathbb{GAL}_0$ \cite{de Saxce 2011, AffineMechBook}:
\begin{equation}
\label{Galilean gravitation eqn}
\Gamma =\left( {{\begin{array}{*{20}c}
 0 \hfill & 0 \hfill \\
 {j\,(\Omega )\,d\,x-g\,d\,t} \hfill & {j\,(\Omega )\,d\,t} \hfill \\
\end{array} }} \right)
\end{equation}
where $g$ is a 3-column collecting the $g^j=-\Gamma _{00}^j $ and identified to the gravity (\cite{Cartan 1923}, \cite{Souriau MMC 123}), while $\Omega $ is a 3-column vector associated by the mapping $j^{-1}$ to the skew-symmetric matrix the elements of which are $\Omega _j^i =\Gamma _{j0}^i $. The spinning vector $\Omega $ can be interpreted as representing Coriolis' effects \cite{Souriau MMC 123, AffineMechBook}. Indeed, for a spinless particle of mass $m$, let the $4$-velocity and linear $4$-momentum be respectively:
\begin{equation}
U = \dot{X} = \left( {{\begin{array}{c}
                       1 \\
                       v \\
       \end{array} }} \right),\qquad 
        T = m\,U = \left( {{\begin{array}{c}
                       m \\
                       p \\
       \end{array} }} \right)\ .
\label{U & T} 
\end{equation}
For particles in the gravitation field, the covariant law of the motion \cite{Cartan 1923}
$$ \nabla_{U} T =  \dot{T} + \Gamma (U)\,T = 0
$$
in a Galilean chart itemizes (\cite{Souriau 1970}, p. 133, formula (12.47) or \cite{Souriau 1997b} for its English translation, \cite{AffineMechBook}, p. 42, formula [3.47]):
\begin{equation}
    \dot{m} = 0,\qquad \dot{p} = m\,( g - 2\, \Omega \times v)
\label{Souriau eqn of motion} 
\end{equation}
The last term of the right hand side is Coriolis' force. It allows explaining Foucault's pendulum without neglecting the centripetal force \cite{AffineMechBook}. The inertial charts are the ones in which the spinning $\Omega$ vanishes. 

Considering a transition map $X' \mapsto X$ between Galilean charts, a Galilean gravitation is modified according to the transformation laws (\cite{Souriau 1970}, p. 138, formula (12.67) or \cite{Souriau 1997b} for its English translation, \cite{AffineMechBook}, p. 43, Theorem 3.2):
\begin{equation}
   \Omega = R\,\Omega' - \omega\ , 
\label{transformation law Omega} 
\end{equation}
\begin{equation}
    g - 2\,\Omega \times v = a_t + R\,(g' - 2\,\Omega' \times v')\ .
\label{transformation law of g - 2 Omega x v} 
\end{equation}   
where the acceleration of transport is:
\begin{equation}
a_t = \dot{u} + \omega \times (v - u)\ .
\label{acceleration of transport} 
\end{equation}

\vspace{.5cm}

\subsection{Potentials of the Galilean gravitation}

The motion of a spinless particle can be described in the tangent bundle $T \mathcal{M}$ by a point of coordinates in a Galilean chart:
$$ \eta  =\left( \begin{array} {c}
                       X \\                      
                       U \\
                    \end {array} \right)
              =\left( \begin{array} {c}
                       t \\
                       x \\
                       1 \\
                       v \\
                    \end {array} \right)
            \in \mathbb{R}^8\ . 
$$
In short notations of exterior calculus, the presymplectic $2$-form reads:
\begin{equation}
   \omega =  m\,\left[(d v_i - g_i\, dt) \wedge (d x_i - v_i\,dt) -  \Omega_{ij}\,dx_i \wedge dx_j \right] \ ,
\label{omega = m ((dv_i - g_i dt) wedge (dx_i - v_i dt) - 2 Omega_ij dx_i wedge dx_j)} 
\end{equation}
where --for sake of easiness-- all the indices are lowered, the standard convention of summation on the repeated indices  is used and $\Omega_{ij}$ is the element at the intersection of the $i$-th row and the $j$-th column of the matrix $j (\Omega)$. Indeed, the condition :
$$ \iota_{d\eta}\,\omega = 0\ ,
$$ 
restitues the equation of motion (\ref{Souriau eqn of motion}). The closure condition $d \omega = 0$ of the presymplectic form gives \cite{AffineMechBook}:
\begin{equation}
   curl\,g + 2 \,\frac{\partial \Omega}{\partial t} = 0,\qquad div\,\Omega = 0\ .
\label{curl g + 2 (partial Omega / partial t) = 0 & div Omega = 0} 
\end{equation}
then there exist potentials $(x,t) \mapsto \phi (x,t) \in \mathbb{R}, (x,t) \mapsto A (x,t) \in \mathbb{R}^3$ such that:
\begin{equation}
    g = - grad\,\phi - \dfrac{\partial A}{\partial t},\qquad  
     \Omega = \frac{1}{2}\,curl\,A\ .
\label{Galilean gravitation potentials} 
\end{equation}
They are defined modulo a gauge:
\begin{equation}
     \phi^{\ast} = \phi - \dfrac{\partial f}{\partial t},\quad  
        A^{\ast} = A + grad\,f\ .
\label{gauge transformation} 
\end{equation}
 The corresponding Lagrangian: 
\begin{equation}
    \mathcal{L}  =  \frac{1}{2}\, m \parallel v \parallel^2 +  m\,A \cdot v - m\,\phi\ ,
\label{L = T - U} 
\end{equation}  
is a Galilean invariant provided:
\begin{equation}
     \phi' = \phi - A \cdot u -\dfrac{1}{2} \parallel u \parallel^2,\qquad
     A' = R^T (A + u)\ .
     \label{phi' & A'} 
\end{equation}
that ensures the conditions (\ref{Galilean gravitation potentials}) are preserved under any galileomorphism (\cite{AffineMechBook}, p. 113, Theorem 6.1.). As it will be seen further, it is worth to remark that $\phi$ is not the usual Newtonian potential. Indeed, this last one has no the Galilean covariance. $\phi$ is a more general object allowing to recover this covariance. For instance, let us consider free particle moving along a straight line at constant velocity (uniform straight motion) in a given Galilean charts $X'$. This situation can be modeled, modulo a gauge, by vanishing potentials:
\begin{equation}
   \phi'  = 0, \qquad A' = 0
\label{phi' = 0 & A' = 0} 
\end{equation}
generating  null accelerations because (\ref{Galilean gravitation potentials}) entails:
\begin{equation}
   g' = \Omega' = 0
\label{g' = Omega' = 0} 
\end{equation}
 Of course in another Galilean chart associated to an observer in an accelerated motion with respect to $X'$, the particle is no more in uniform straight motion. Combining (\ref{phi' & A'}) and (\ref{phi' = 0 & A' = 0}) leads to:
\begin{equation}
   \phi = - \dfrac{1}{2} \parallel u \parallel^2 , \qquad A = - u
\label{phi = - (1/2) norm(u)^2 & A = - u} 
\end{equation}
that, owing to (\ref{transformation law of g - 2 Omega x v}) and (\ref{g' = Omega' = 0}), generates the so-called inertial forces:
\begin{equation}
   m (g - 2\,\Omega \times v) = m\,a_t 
\label{inertial forces} 
\end{equation}

\section{Galilean tensors} 
\label{Section Galilean tensors} 

By restriction of the transformation law of tensors to a subgroup $G \subset \mathbb{GL} (n)$, we obtain the $G$-tensors. For instance, Euclidean tensors are $\mathbb{SO} (3)$-tensors. The $\mathbb{GAL}_0$-tensors are called Galilean tensors. $G$-tensors may be seen as orbits for the action of $G$ onto the tensor components. Examples of Galilean tensors are:
\begin{itemize}
\item Vectors: $V^{\alpha'} = (P^{-1})^{\alpha'}_\beta V^\beta$ or $V' = P^{-1} V$ in matrix form with $P\in \mathbb{GAL}_0$ applied to (\ref{U & T}) gives the composition formula of Galilean velocities: 
\begin{equation}
    v' = R^T (v - u)\ .
\label{v'=R^T (v - u)} 
\end{equation}
In particular $v = u + v'$ is the additive decomposition of velocities in classical mechanics. Owing to (\ref{phi' & A'}), the column:
\begin{equation}
    C  = \left( {{\begin{array}{c}
                       1 \\
                       - A \\
       \end{array} }} \right)\ ,
\label{C = (1 -A)} 
\end{equation}
represents a Galilean vector $\bm{C}$ that could be called Coriolis' vector because it generates the corresponding force.
\item Covectors: $\tau_{\alpha'} =\tau^\beta P^\beta_{\alpha'}$ or $\tau' = \tau\,P$
applied to the covector $\bm{\tau}$ of the Toupinian structure shows that it is represented by an invariant $4$-row $  \tau = (1, 0, 0,  0)$.
\item $2$-covariant tensors:  $T_{\alpha'\beta'} = P^\mu_{\alpha'} P^\nu_{\beta'} T_{\mu\nu}$ 
or $T' = P^T T \,P$ applied to the symmetric tensor:
$$   T = \left( {{\begin{array}{cc}
                       a & w^T \\
                       w & M     \\
       \end{array} }} \right)\ ,
$$ 
gives: 
\begin{eqnarray}
     a' & = & a + 2\,w\cdot u + u\,(M\,u),\nonumber\\
     w' & = & R^T (w + M\,u),\qquad 
     M' = R^T M\,R\ .
\label{transformation law of a & w & M} 
\end{eqnarray}
For instance, taking into account (\ref{phi' & A'}), the tensor represented by:
\begin{equation}
   \overset{(0)}{G} = \left( {{\begin{array}{cc}
                       2\,\phi & - A^T \\
                       - A & - 1_{\mathbb{R}^3}    \\
       \end{array} }} \right)\ , 
\label{G = matrix ( - 2 phi A^ T A 1_R^3 )}
\end{equation}
is a Galilean $2$-covariant tensor. It has only one independant Galilean invariant (\cite{AffineMechBook}, Theorem 6.1.):
\begin{equation}
   I_0 = - \frac{1}{2}\, \det (\overset{(0)}{G}) =  \phi\; + \parallel A \parallel^2 /\,2\ .
\label{galilean invariant of phi & A} 
\end{equation}
\item $2$-contravariant tensors:  $T^{\alpha'\beta'} = (P^{-1})_\mu^{\alpha'} (P^{-1})_\nu^{\beta'} T^{\mu\nu}$ 
or $T' = P^{-1} T \,P^{-T}$ applied to the symmetric tensor:
$$   T = \left( {{\begin{array}{cc}
                       b & v^T \\
                       v & N     \\
       \end{array} }} \right)\ ,
$$ 
gives: 
$$ b'  =  b,\qquad  v' = R^T (v -  b\,u)\,
$$
$$ N'  =  R^T (N - u\,v^T - v\,u^T - b\,u\,u^T)\,R\ .
$$
For instance, the tensor $\bm{h}$ of the Toupinian structure is a Galilean $2$-contravariant tensor represented by the invariant matrix:
$$   h = \left( {{\begin{array}{cc}
                       0 & 0^T \\
                       0 & 1_{\mathbb{R}^3}    \\
       \end{array} }} \right)\ , 
$$
\end{itemize}

\section{Space-time metrics}
\label{Section Space-time metrics}

In the approximation of a weak field of Galilean gravitation, we claim that the space-time metrics reads in Galilean coordinate systems:
\begin{equation}
   G =\left( {{\begin{array}{cc}
               c^2 + 2\,\phi \hfill &   -A^T             \hfill \\
               - A           \hfill & - 1_{\mathbb{R}^3} \hfill \\
   \end{array} }} \right)\ ,
\label{Weak field metrics} 
\end{equation}
where $c$ is the speed of the light. Assuming it is finite but huge, we introduce the small parameter:
$$ \epsilon = c^{-2}\ .
$$
Hence the metric is expanded as:
$$  G = \epsilon^{-1}\,\overset{(-1)}{G} + \overset{(0)}{G}
      = \epsilon^{-1}\,\left( {{\begin{array}{cc}
                1 \hfill & 0 \hfill \\
                0 \hfill & 0 \hfill \\
   \end{array} }} \right)
   + \left( {{\begin{array}{cc}
               2\,\phi \hfill &   -A^T             \hfill \\
               - A     \hfill & - 1_{\mathbb{R}^3} \hfill \\
   \end{array} }} \right)\ .
$$
The former term represents $\epsilon^{-1} = c^2$ multiplied by the tensor product :
$$ \overset{(-1)}{\bm{G}} = \bm{\tau} \otimes \bm{\tau}\ ,
$$  
of the covector of the Toupinian structure by itself then it is dominant. The latter one represents the symmetric $2$-covariant tensor (\ref{G = matrix ( - 2 phi A^ T A 1_R^3 )})  constructed from the Galilean gravitation potentials $\phi$ and $A$. 
Hence both terms are separately symmetric $2$-covariant Galilean tensors, allowing to write the coordinate-free expansion:
$$  \bm{G} = \epsilon^{-1}\,\overset{(-1)}{\bm{G}} + \overset{(0)}{\bm{G}}\ .
$$
Moreover, we recover Minkowski's metrics if the gravitation potentials vanish. By Frobenius formula, we calculate its determinant:
$$ \det \,G = - (\epsilon^{-1} + 2\,I_0)\ ,
$$
where $I_0$ is the Galilean invariant (\ref{galilean invariant of phi & A}). In the weak field approximation, $\epsilon^{-1}$ is dominant, thus we may assume that it is negative (hence non vanishing and (\ref{Weak field metrics}) represents a metrics). In the sequel, we need the expansion:
\begin{equation}
   \sqrt{ - \det \,G}\cong c\, \left(1 + \epsilon\,I_0\right)
                     = c\, \left(1 + \epsilon\,\left(\phi 
                                               + \frac{1}{2}\,\parallel A \parallel^2\right)
                            \right)\ .
\label{sqrt( - det G) = c (1 + epsilon (phi + (1/2) norm(A)^2))} 
\end{equation}
By Frobenius formula and with the expansion:
$$ (\epsilon^{-1} + 2\,I_0)^{- 1} \cong \epsilon^{-1} - 2\,I_0\ ,
$$
we obtain the approximation of the inverse of Gram's matrix:
\begin{equation}
  G^{-1} \cong \left( {{\begin{array}{cc}
                0 \hfill & 0                  \hfill \\
                0 \hfill & - 1_{\mathbb{R}^3} \hfill \\
   \end{array} }} \right)
   + \epsilon\,
     \left( {{\begin{array}{cc}
               1   \hfill &   -A^T \hfill \\
               - A \hfill & A\,A^T \hfill \\
   \end{array} }} \right)
   - 2\,\epsilon^2 I_0\,
     \left( {{\begin{array}{cc}
               1   \hfill &   -A^T \hfill \\
               - A \hfill & A\,A^T \hfill \\
   \end{array} }} \right)\ .
 \label{Weak field contravariant metrics expansion} 
\end{equation}
The former term is a Galilean $2$-contravariant tensor, in fact the opposite of the tensor $\bm{h}$ of the Toupinian structure. Applying the transformation law (\ref{transformation law of a & w & M}) with (\ref{phi' & A'}), it is straightforward to verify that the matrix:
$$   \left( {{\begin{array}{cc}
               1   \hfill &   -A^T \hfill \\
               - A \hfill & A\,A^T \hfill \\
   \end{array} }} \right)\ ,
$$
represents the Galilean $2$-contravariant tensor $\bm{C} \otimes \bm{C} $, tensor product of Coriolis' vector by itself. As $I_0$ is a Galilean invariant, each term of the  expansion of $G^{-1}$ represents a Galilean $2$-contravariant tensor, allowing to write the expansion:
$$  \bm{G}^{-1} \cong \overset{(0)\quad}{\bm{G}^{-1}} 
                  + \epsilon\,\overset{(1)\quad}{\bm{G}^{-1}}
                  + \epsilon^2\overset{(2)\quad}{\bm{G}^{-1}}\ .
$$
We verify that for $G$ given by (\ref{Weak field metrics}) and $G^{-1}$ given by (\ref{Weak field contravariant metrics expansion}):
$$ G\,G^{-1} \cong 1_{\mathbb{R}^4} + O (\epsilon^2)\ .
$$
In the sequel, the expansion (\ref{Weak field contravariant metrics expansion}) will be troncated to the two former terms:
\begin{equation}
  G^{-1} \cong \left( {{\begin{array}{cc}
                0 \hfill & 0                  \hfill \\
                0 \hfill & - 1_{\mathbb{R}^3} \hfill \\
   \end{array} }} \right)
   + \epsilon\,
     \left( {{\begin{array}{cc}
               1   \hfill &   -A^T \hfill \\
               - A \hfill & A\,A^T \hfill \\
   \end{array} }} \right)\ .
 \label{Weak field contravariant metrics expansion troncated} 
\end{equation}

\section{Functional due to the matter}
\label{Section Functional due to the matter} 

In \cite{Souriau 1964}, the motion of a continuum is described by a line bundle $\pi_0: \mathcal{M}\mapsto\mathcal{M}_0$ where $\mathcal{M}_0$ is a manifold of dimension 3 representing the matter and each fiber is the trajectory of a material particle identified by its reference position $\bm{x}_0 = \pi_0 (\bm{X})$. In local charts, it is represented by $ x_0 \in \mathbb{R}^3$ (its position at a given date) and its motion is determined thanks to a mapping $ (t,x) \mapsto x_0 = \kappa (t,x) $ which identifies the material point located at position $x$ at time $t$. The space-time coordinates $(t,x)$ are Eulerian while $(t,x_0)$ are Lagrangian. As obviously $x_0$ is an invariant of the motion, the material derivative vanishes:
$$   \frac{dx_0}{dt} = \frac{\partial x_0}{\partial t} + \frac{\partial x_0}{\partial x} \frac{dx}{dt} = 0\ .
$$
Then, using usual notation for the deformation gradient :
$$ F = \frac{\partial x}{\partial x_0} \ ,
$$
the spacetime gradient is the $3 \times 4$ matrix:
\begin{equation}
     \frac{\partial x_0}{\partial X} = \left(\frac{\partial x_0}{\partial t}, \frac{\partial x_0}{\partial x}   \right)
                                                  = (- F^{-1} v, F^{-1})\ .
\label{partial s' / partial X = (- F^(-1) v, F^(-1))} 
\end{equation}
In Hilbert-Einstein variational theory of the General Relativity, the functional is compound of a term due to the space-time curvature and a term due to the presence of the matter:
\begin{equation}
   \int \,(p_M + p_G)\sqrt{- \det G}\,d^4 X
\label{Hilbert-Einstein functional} 
\end{equation}
To build the latter one, we introduce, according to \cite{Souriau 1964} (Formula (39.10), page 375), the conformation tensor represented by:
$$ \mathcal{D} = - \frac{\partial x_0}{\partial X} \left( \frac{\partial x_0}{\partial X}\right)^*\ ,
$$ 
where the adjoint matrix is given by:
$$ \left( \frac{\partial x_0}{\partial X}\right)^* = 
            G^{-1} \left( \frac{\partial x_0}{\partial X}\right)^T\ ,
$$
the $3D$ space metrics being represented by the identity matrix in Galilean coordinate systems. Taking into account (\ref{partial s' / partial X = (- F^(-1) v, F^(-1))}) and (\ref{Weak field contravariant metrics expansion}), we obtain the approximation:
$$ \mathcal{D} \cong F^{-1} \left( 1_{\mathbb{R}^3} - \epsilon (v + A)\,(v + A)^T \right) F^{-T} + O(\epsilon^2)\ .
$$
and the one of its determinant:
$$ \sqrt{\det \mathcal{D}} = \sqrt{(\det (F^T F))^{-1}\,(1 - \epsilon\,\parallel v + A\parallel^2)}
$$
\begin{equation}
 \sqrt{\det \mathcal{D}}  \cong (\det F)^{-1} \left(1 - \frac{\epsilon}{2}\,\parallel v + A\parallel^2\right)\ .
\label{sqrt(det D) = 1 - (epsilon/2) norm(v + A)^2} 
\end{equation}
According to \cite{Souriau 1964} (Formula (39.11), page 375), the term of the functional due to the matter is obtained by integration over the space-time of:
$$ p_M \sqrt{- \det G} =  c^2 \rho_0 (x_0)\,\sqrt{\det \mathcal{D}}\,\sqrt{- \det G}\ .
$$
Taking into account the law of variation of the density with respect to the deformation gradient:
$$ \rho = \frac{\rho_0 (x_0)}{\det F}\ ,
$$
and  (\ref{sqrt( - det G) = c (1 + epsilon (phi + (1/2) norm(A)^2))}) and (\ref{sqrt(det D) = 1 - (epsilon/2) norm(v + A)^2}), its approximation reads:
\begin{equation}
p_M \sqrt{- \det G} \cong c\left[  \rho\,c^2 
                - \rho\,\left(\dfrac{1}{2}\,\parallel v \parallel^2 + A \cdot v - \phi\right)\right]  \ .
\label{p_M sqrt(- G) cong } 
\end{equation}
In the bracket, the former term is dominant and represents the energy of the mass at rest (by volume unit). In the latter one, we recognize the Galilean Lagrangian (\ref{L = T - U}) of the matter in absence of internal energy. Then every term of the expansion of  (\ref{p_M sqrt(- G) cong }) is a Galilean invariant.

\section{The connection}
\label{Section The connection} 

The Galilean connection (\ref{Galilean gravitation eqn}) satisfies automatically the compatibility conditions with the Toupinian structure tensors:
$$ \nabla\, \bm{\tau} = \bm{0},\qquad 
    \nabla\, \bm{h} = \bm{0}
$$
Then it is not uniquely determined by the Toupinian structure. The idea is to derive the connection from the potentials $\phi$ and $A$ through the metrics of Section \ref{Section Space-time metrics}, using standard Levi-Civita connection. To construct the term of Hilbert-Einstein principle due to the space-time curvature, we need to calculate successively Christoffel’s symbols of the connection, the corresponding Riemann-Christoffel tensor, Ricci tensor and scalar curvature. For the metrics (\ref{Weak field metrics}), Christoffel's symbols of the first kind read:
$$    \left[00, 0 \right] =   \frac{\partial \phi}{\partial t},\qquad 
   \left[00, i \right] = - \frac{\partial \phi}{\partial x^i} - \frac{\partial A_i}{\partial t},\qquad
   \left[0i, 0 \right] =   \frac{\partial \phi}{\partial x^i}\ , 
$$
$$ \left[0i, j \right] = \frac{1}{2}\,\left(   \frac{\partial A_i}{\partial x^j}
                                             - \frac{\partial A_j}{\partial x^i}\right),\qquad
   \left[ij, 0 \right] = \frac{1}{2}\,\left(   \frac{\partial A_i}{\partial x^j}
                                             + \frac{\partial A_j}{\partial x^i}\right)\ .
$$
of which we derive Christoffel’s symbols of the second with the troncated expansion (\ref{Weak field contravariant metrics expansion troncated}):
$$ \Gamma^\mu_{\alpha\beta} =            \overset{(0)\quad}{\Gamma^{\mu}_{\alpha\beta}}
                             + \epsilon\,\overset{(1)\quad}{\Gamma^{\mu}_{\alpha\beta}}
                             + O (\epsilon^2)\ , 
$$
Taking into account (\ref{Galilean gravitation potentials}), we recognize at the order zero the Galilean gravitation (\ref{Galilean gravitation eqn}) of non vanishing terms:
$$ \overset{(0)\quad}{\Gamma^{i}_{00}} = - g^i,\qquad
   \overset{(0)\quad}{\Gamma^{j}_{i0}} =   \Omega^j_i\ .
$$
At the first order, we obtain:
\begin{equation}
   \overset{(1)\quad}{\Gamma^{0}_{00}} =   \frac{\partial \phi}{\partial t} - A \cdot g,\qquad
   \overset{(1)\quad}{\Gamma^{i}_{00}} = - \frac{\partial \phi}{\partial t}\,A_i\ ,
\label{Christoffel symbols term order 1 part 1} 
\end{equation}
\begin{equation}
   \overset{(1)\quad}{\Gamma^{0}_{0i}} 
 = \overset{(1)\quad}{\Gamma^{0}_{i0}} = \frac{\partial \phi}{\partial x^i} 
                - \frac{1}{2}\,\left(\frac{\partial A_i}{\partial x^j} - \frac{\partial A_j}{\partial x^i} \right)\,A_j                   
                   = \left(grad\,\phi - \Omega\times A \right)^i\ ,
\label{Christoffel symbols term order 1 part 2} 
\end{equation}
\begin{equation}
   \overset{(1)\quad}{\Gamma^{j}_{i0}} = - A^j \left(\frac{\partial \phi}{\partial x^i}
                                            + \Omega^r_i A_r\right)\ ,  
\label{Christoffel symbols term order 1 part 3} 
\end{equation}
\begingroup\makeatletter\def\f@size{10}\check@mathfonts
\begin{equation}
   \overset{(1)\quad}{\Gamma^{0}_{ij}} = - \frac{1}{2}\,\left(\frac{\partial A_i}{\partial x^j} 
                                         + \frac{\partial A_j}{\partial x^i} \right)
                                       = - \left( grad_s A \right)^i_j,\quad
                                       \overset{(1)\quad}{\Gamma^{k}_{ij}}  = A^k\,\left( grad_s A \right)^i_j\ ,
\label{Christoffel symbols term order 1 part 4} 
\end{equation} \endgroup
where --for sake of easiness-- the index of $A$ is equally lowered or raised. Also for tight calculations in the sequel, we introduce simplified notations :
\begin{equation}
   a = \overset{(1)\quad}{\Gamma^{0}_{00}} =   \frac{\partial \phi}{\partial t} - A \cdot g\ ,
\label{a = partial phi / partial t - A cdot g} 
\end{equation}
\begin{equation}
   B_i = \overset{(1)\quad}{\Gamma^{0}_{0i}} 
   = \overset{(1)\quad}{\Gamma^{0}_{i0}}  = \delta_{ik} \left(grad\,\phi - \Omega\times A \right)^k\ ,
\label{B_i = } 
\end{equation}
\begin{equation}
   D_{ij} =   \overset{(1)\quad}{\Gamma^{0}_{ij}}  = - \delta_{ik} \left( grad_s A \right)^k_j\ ,
\label{D_(ij) = } 
\end{equation}
hence :
$$ \overset{(1)\quad}{\Gamma^{j}_{i0}}  = A^j B_i = (A\,B^T)^j_i,\qquad 
   \overset{(1)\quad}{\Gamma^{k}_{ij}}  = A^k D_{ij}\ . 
$$

Comparing to Christoffel’s symbols of the Bargmannian gravitation (\cite{AffineMechBook}, page 273, formulae [13.75] and [13.76]), it is worth to observe that:
$$ \overset{(1)\quad}{\Gamma^0_{\alpha\beta}}  = \Gamma^4_{\alpha\beta}\ , 
$$
where the index 4 correspond to the extra dimension of the Bargmannian representation.

\section{Riemann-Christoffel tensor}
\label{Section Riemann-Christoffel tensor} 

The Riemann-Christoffel tensor can be expanded as:
$$ R^\delta_{\alpha\beta\gamma} =        \overset{(0)\quad}{R^\delta_{\alpha\beta\gamma}}
                             + \epsilon\,\overset{(1)\quad}{R^\delta_{\alpha\beta\gamma}}
                             + O (\epsilon^2)\ , 
$$
with at the order zero:
$$    \overset{(0)\quad}{R^\delta_{\alpha\beta\gamma}} = 
   \overset{(0)\quad}{\Gamma^\delta_{\alpha\mu}} \overset{(0)\quad}{\Gamma^\mu_{\beta\gamma} }
                 - \overset{(0)\quad}{\Gamma^\delta_{\beta\mu}} \overset{(0)\quad}{\Gamma^\mu_{\alpha\gamma}}
                 + \frac{\partial \overset{(0)\quad}{\Gamma^\delta_{\beta\gamma}}}{\partial X^\alpha}
                 - \frac{\partial \overset{(0)\quad}{\Gamma^\delta_{\alpha\gamma}}}{\partial X^\beta}\ .
$$
The only non vanishing components are:
\begin{equation}
   \overset{(0)\quad}{R^i_{0j0}} = - \overset{(0)\quad}{R^i_{j00}} =   \frac{\partial g^i} {\partial x^j} 
                             + \frac{\partial \Omega^i_j} {\partial t}
                             + \Omega^i_k\, \Omega^k_j ,
\label{R^i_(0j0) = - R^ i_(j00) = BIS} 
\end{equation}
\begin{equation}
\overset{(0)\quad}{R^i_{kj0}} = - \overset{(0)\quad}{R^i_{jk0}} =   \frac{\partial \Omega^i_j} {\partial x^k} 
                             - \frac{\partial \Omega^i_k} {\partial x^j},\qquad
   \overset{(0)\quad}{R^i_{k0j}} = - \overset{(0)\quad}{R^i_{0kj}} =  \frac{\partial \Omega^i_j} {\partial x^k}\ .
\label{R^i_(kj0) = - R^ i_(jk0) = & R^i_(k0j) = - R^ i_(0kj) = BIS} 
\end{equation}
At the first order, one has:
$$    \overset{(1)\quad}{R^\delta_{\alpha\beta\gamma}} = 
   \overset{(0)\quad}{\Gamma^\delta_{\alpha\mu}} \overset{(1)\quad}{\Gamma^\mu_{\beta\gamma} }
 + \overset{(1)\quad}{\Gamma^\delta_{\alpha\mu}} \overset{(0)\quad}{\Gamma^\mu_{\beta\gamma} }
                 - \overset{(0)\quad}{\Gamma^\delta_{\beta\mu}} \overset{(1)\quad}{\Gamma^\mu_{\alpha\gamma}}
                 - \overset{(1)\quad}{\Gamma^\delta_{\beta\mu}} \overset{(0)\quad}{\Gamma^\mu_{\alpha\gamma}}
                 + \frac{\partial \overset{(1)\quad}{\Gamma^\delta_{\beta\gamma}}}{\partial X^\alpha}
                 - \frac{\partial \overset{(1)\quad}{\Gamma^\delta_{\alpha\gamma}}}{\partial X^\beta}\ .
$$
Since $\overset{(1)\quad}{R^\delta_{\alpha\beta\gamma}}$ is skew-symmetric in $\alpha$ and $\beta$, we shall calculate only the terms with $\alpha < \beta$. Taking into account (\ref{Christoffel symbols term order 1 part 1}) to (\ref{Christoffel symbols term order 1 part 3}) and the tight notations (\ref{a = partial phi / partial t - A cdot g}) to (\ref{D_(ij) = }), one has:
\begin{equation}
 \overset{(1)\quad}{R^0_{0l0}} = B_i \Omega^i_l - D_{li} g^i 
                                 + \dfrac{\partial B_l}{\partial t} 
                                 - \dfrac{\partial a}{\partial x^l} 
\label{1st order R^ 0_(0l0) = } 
\end{equation}
\begin{equation}
 \overset{(1)\quad}{R^0_{kl0}} = D_{lj} \Omega^j_k - D_{kj} \Omega^j_l
                                 + \dfrac{\partial B_l}{\partial x^k} 
                                 - \dfrac{\partial B_k}{\partial x^l} 
\label{1st order R^ 0_(kl0) = } 
\end{equation}
\begin{equation}
 \overset{(1)\quad}{R^0_{0lm}} = D_{lj} \Omega^j_m - \dfrac{\partial D_{lm}}{\partial t}
                                 - \dfrac{\partial B_m}{\partial x^l} 
\label{1st order R^ 0_(0lm) = } 
\end{equation}
\begin{equation}
 \overset{(1)\quad}{R^0_{klm}} =  \dfrac{\partial D_{km}}{\partial x^l}
                                  -   \dfrac{\partial D_{lm}}{\partial x^k}                            
\label{1st order R^ 0_(klm) = } 
\end{equation}
\begingroup\makeatletter\def\f@size{10}\check@mathfonts
\begin{equation}
 \overset{(1)\quad}{R^j_{0l0}} = - g^j B_l - \Omega^j_i A^i B_l - \Omega^i_l A^j B_i 
                                  - a\, \Omega^j_l + A^j D_{li} g^ i 
                                 - \dfrac{\partial}{\partial t} (A^j B_l)
                                 + \dfrac{\partial}{\partial x^l} 
                                  \left(\dfrac{\partial \phi}{\partial t}\,A^j \right) 
\label{1st order R^ j_(0l0) = } 
\end{equation} \endgroup
\begin{equation}
 \overset{(1)\quad}{R^j_{kl0}} = \Omega^j_k B_l - \Omega^j_l B_k 
                                   + A^j ( D_{ki} \Omega^i_l - D_{li} \Omega^i_k ) 
                                   -  \dfrac{\partial}{\partial x^k} (A^j B_l)
                                   +  \dfrac{\partial}{\partial x^l} (A^j B_k)                           
\label{1st order R^ j_(kl0) = } 
\end{equation}
\begin{equation}
 \overset{(1)\quad}{R^j_{0lm}} = A^i D_{lm} \Omega^j_i - A^j D_{li} \Omega^i_m 
                                  - B_m \Omega^j_l
                                  + \dfrac{\partial}{\partial t} (A^j D_{lm})
                                  + \dfrac{\partial}{\partial x^l} (A^j B_m) 
\label{1st order R^ j_(0lm) = } 
\end{equation}
\begin{equation}
 \overset{(1)\quad}{R^j_{klm}} = D_{km} \Omega^j_l - D_{lm} \Omega^j_k 
                                  + \dfrac{\partial}{\partial x^k} (A^j D_{lm})
                                  - \dfrac{\partial}{\partial x^l} (A^j D_{km})
\label{1st order R^ j_(klm) = } 
\end{equation}

\section{Ricci tensor}
\label{Section Ricci tensor} 

Next, the components of the Ricci tensor $\bm{R}'$ are deduced from the ones of the Riemann-Christoffel tensor by contraction:
$$ R'_{\beta\gamma} = R^\alpha_{\alpha\beta\gamma}\ ,
$$
that leads to the expansion:
\begin{equation}
   R'_{\beta\gamma} =        \overset{(0)\quad}{R'_{\beta\gamma}}
                             + \epsilon\,\overset{(1)\quad}{R'_{\beta\gamma}}
                             + O (\epsilon^2)\ , 
\label{expansion of Ricci tensor troncated} 
\end{equation}
with:
$$ \overset{(i)\quad}{R'_{\beta\gamma}} = \overset{(i)\quad}{R^\alpha_{\alpha\beta\gamma}}\ .
$$
In particular, for the order zero, the only non vanishing components are, taking into account (\ref{R^i_(0j0) = - R^ i_(j00) = BIS}) and (\ref{R^i_(kj0) = - R^ i_(jk0) = & R^i_(k0j) = - R^ i_(0kj) = BIS}):
$$ \overset{(0)\quad}{R'_{00}}
   = - \overset{(0)\quad}{R^i_{0i0}} 
   = - \left( \frac{\partial g^i} {\partial x^i} + \Omega^i_k\, \Omega^k_i \right) \ ,
$$
$$ \overset{(0)\quad}{R'_{0m}}
   = - \overset{(0)\quad}{R^i_{i0m}} 
   =   \frac{\partial \Omega^i_m} {\partial x^i},\qquad
    \overset{(0)\quad}{R'_{m0}}
   = - \overset{(0)\quad}{R^i_{im0}} 
   =   \frac{\partial \Omega^i_m} {\partial x^i} \ .
$$
In a matrix form, the zero order term of the expansion reads:
\begin{equation}
   \overset{(0)}{R'}  = \left( {{\begin{array}{*{20}c}
              2 \parallel \Omega \parallel^2 - div\,g \hfill & (curl\,\Omega)^T \hfill \\
              curl\,\Omega                            \hfill & 0                \hfill \\
            \end{array} }} \right)\ .
\label{0 order R'} 
\end{equation}

For the first order, we obtain, taking into account (\ref{1st order R^ 0_(0lm) = }), (\ref{1st order R^ j_(0l0) = }), (\ref{1st order R^ j_(0lm) = }), (\ref{1st order R^ j_(klm) = }) and the skew-symmetry of $\Omega$:
$$  \overset{(1)\quad}{R'_{00}} = \overset{(1)\quad}{R^0_{000}} + \overset{(1)\quad}{R^j_{j00}}
                                 = - \overset{(1)\quad}{R^j_{0j0}}
$$
$$ \overset{(1)\quad}{R'_{00}} = - A^j D_{ji} g^ i 
                                  + \dfrac{\partial}{\partial t} (A^j B_j)
                                  - \dfrac{\partial}{\partial x^j} 
                                    \left(\dfrac{\partial \phi}{\partial t}\,A^j \right) 
                                    + g^i B_i + 2\, \Omega^j_i A^i B_j
$$
$$  \overset{(1)\quad}{R'_{0m}} = - \overset{(1)\quad}{R^j_{0jm}}
$$
\begin{equation}
 \overset{(1)\quad}{R'_{0m}} = - \Omega^j_i A^i D_{jm} + A^j D_{ji} \Omega^i_m 
                                - \dfrac{\partial}{\partial t} 
                                  \left(A^j D_{jm} \right) 
                                - \dfrac{\partial}{\partial x^j} (A^j B_m) - D_{mi} g^i
\label{1st order R'_(0m) = } 
\end{equation}
$$ \overset{(1)\quad}{R'_{m0}} = \overset{(1)\quad}{R^0_{0m0}} + \overset{(1)\quad}{R^j_{jm0}} 
$$
\begin{eqnarray}
\overset{(1)\quad}{R'_{m0}}  & = & 
                                  \dfrac{\partial B_m}{\partial t} 
                                 - \dfrac{\partial a}{\partial x^m} 
                                + A^j ( D_{ji} \Omega^i_m - D_{mi} \Omega^i_j )\nonumber\\ 
                           & &    -  \dfrac{\partial}{\partial x^j} (A^j B_m)
                                   +  \dfrac{\partial}{\partial x^m} (A^j B_j)
                                   - D_{mi} g^i 
\label{1st order R'_(m0) = } 
\end{eqnarray}
$$ \overset{(1)\quad}{R'_{lm}} = \overset{(1)\quad}{R^0_{0lm}} + \overset{(1)\quad}{R^j_{jlm}} 
$$
\begin{equation}
\overset{(1)\quad}{R'_{lm}}   = D_{lj} \Omega^j_m - \dfrac{\partial D_{lm}}{\partial t}
                                 - \dfrac{\partial B_m}{\partial x^l} 
                                 + D_{jm} \Omega^j_l 
                                  + \dfrac{\partial}{\partial x^j} (A^j D_{lm})
                                  - \dfrac{\partial}{\partial x^l} (A^j D_{jm})
\label{1st order R'_(lm) = } 
\end{equation}

For Levi-Civita connection, Ricci tensor is symmetric, hence because of (\ref{expansion of Ricci tensor troncated}):
$$ \overset{(0)\quad}{R'_{\beta\gamma}}
                             + \epsilon\,\overset{(1)\quad}{R'_{\beta\gamma}}
                             + O (\epsilon^2) 
 = \overset{(0)\quad}{R'_{\gamma\beta}}
                             + \epsilon\,\overset{(1)\quad}{R'_{\gamma\beta}}
                             + O (\epsilon^2)
$$
As the expansion is unique, each term of the expansion must be symmetric:
$$ \overset{(0)\quad}{R'_{\beta\gamma}} = \overset{(0)\quad}{R'_{\gamma\beta}},\qquad
   \overset{(1)\quad}{R'_{\beta\gamma}} = \overset{(1)\quad}{R'_{\gamma\beta}}\ .
$$
It is so for the zero order term as it can be seen in (\ref{0 order R'}). Let us check it for the term of the first order.

First, we want to verify:
$$ \overset{(1)\quad}{R'_{lm}} = \overset{(1)\quad}{R'_{ml}} \ .
$$ 
The second and fifth terms of (\ref{1st order R'_(lm) = }) are symmetric in $l$ and $m$. The sum of the first and fourth ones so is too. It remains to check whether:
$$ \dfrac{\partial B_m}{\partial x^l} + \dfrac{\partial}{\partial x^l} (A^j D_{jm})\ ,
$$
is symmetric in $l$ and $m$, or equivalently whether the matrix:
\begin{equation}
   M = \dfrac{\partial}{\partial x} (B + D\,A)\ ,
\label{M = (partial / partial x) (B + D A)} 
\end{equation}
is symmetric. 
Owing to (\ref{Galilean gravitation potentials}) and (\ref{D_(ij) = }), one has:
\begin{equation}
   grad\,A = D - j (\Omega)\ ,
\label{grad A = D - j (Omega)} 
\end{equation}
hence:
$$ B + D\,A = grad\,\phi + (D - j (\Omega))\, A = grad\,\phi + (grad\,A)\,A\ ,
$$
and using:
$$    grad\,(u\cdot v) = (grad\,u)\,v + (grad\,v)\,u\ ,
$$
one has:
$$ B + D\,A = grad\,\left(\phi + \frac{1}{2}\,\parallel A \parallel^2\right) = grad\,I_0\ ,
$$
where $I_0$ is the Galilean invariant (\ref{galilean invariant of phi & A}). Hence, the matrix (\ref{M = (partial / partial x) (B + D A)}) is symmetric as the Hessian matrix of $I_0$:
$$ M =  \dfrac{\partial}{\partial x} (grad\,I_0)\ .
$$
Next, we want to verify:
\begin{equation}
    \overset{(1)\quad}{R'_{0m}}  = \overset{(1)\quad}{R'_{m0}} \ .
\label{1st order R'_(0m) = R'_(m0)} 
\end{equation}

Firstly, let us remark that $\overset{(1)\quad}{R'_{m0}}$ is the $m$-th component of the $3$-column:
$$ \overset{(1)}{r}_* = (\Omega\,D - D\,\Omega)\,A + \dfrac{\partial B}{\partial t}
                         - grad\,a
                         - div\,(A\,B^T) + grad\,(A \cdot B)
                         - D\,g \ ,
$$
and $\overset{(1)\quad}{R'_{0m}}$ is the $m$-th component of the $3$-column:
$$ \overset{(1)}{r}  = (\Omega\,D - D\,\Omega)\,A 
                                   - \dfrac{\partial}{\partial t} (D\,A) 
                                   - div\,(A\,B^T)
                                   - D\,g\ ,
$$
The symmetry condition is satisfied if the two latter columns are identical, that is when:
$$ \dfrac{\partial}{\partial t} (D\,A)  =  - \dfrac{\partial B}{\partial t} + grad\,a - grad\,(A \cdot B)\ .
$$
or, owing to (\ref{a = partial phi / partial t - A cdot g}) and (\ref{B_i = }):
$$ \dfrac{\partial}{\partial t} (D - j (\Omega))\,A) 
     = - grad\,(A \cdot (g + grad\,\phi)) - grad\,(A \cdot (\Omega \times A))\ ,
$$
where the scalar triple product in the last term vanishes. Taking into account (\ref{Galilean gravitation potentials}) and (\ref{grad A = D - j (Omega)}), one has:
$$ \dfrac{\partial}{\partial t} ((grad\,A)\,A) 
     =   grad \left(A \cdot \dfrac{\partial A}{\partial t} \right)\ ,
$$
$$ \dfrac{\partial}{\partial t} grad \left(\frac{1}{2}\,\parallel A \parallel^2\right) 
     =   grad\,\dfrac{\partial}{\partial t} \left(\frac{1}{2}\,\parallel A \parallel^2\right) \ ,
$$
that proves (\ref{1st order R'_(0m) = R'_(m0)}).

In a matrix form, the term of order one reads:
\begin{equation}
   \overset{(1)}{R'}  = \left( {{\begin{array}{*{20}c}
              \overset{(1)}{r_0} \hfill & \overset{(1)}{r}^T_* \hfill \\
              \overset{(1)}{r}_*  \hfill & \overset{(1)}{R_s}    \hfill \\
            \end{array} }} \right)\ ,
\label{1st order R'} 
\end{equation}
where $\overset{(1)}{R_s}$ is the $3 \times 3$ matrix of elements $\overset{(1)\quad}{R'_{lm}}$.

\section{Scalar curvature}
\label{Section Scalar curvature} 

It is defined as:
$$ R = G^{\alpha\beta} R'_{\alpha\beta} = Tr\,(\bm{G}^{-1} \bm{R}') = Tr\,(G^{-1} R')
$$
owing to the expansion (\ref{Weak field contravariant metrics expansion troncated}) of the contravariant metrics, (\ref{expansion of Ricci tensor troncated}) of Ricci tensor, (\ref{0 order R'}) and (\ref{1st order R'}):
\begingroup\makeatletter\def\f@size{9}\check@mathfonts
$$ G^{-1} R' =  \left( {{\begin{array}{*{20}c}
              0              \hfill & 0 \hfill \\
              - curl\,\Omega \hfill & 0 \hfill \\
            \end{array} }} \right)
            + \epsilon \left( {{\begin{array}{*{20}c}
              J_*     \hfill & (curl\,\Omega)^T      \hfill \\
              - J_* A \hfill & - A\,(curl\,\Omega)^T \hfill \\
            \end{array} }} \right)
            + \epsilon \left( {{\begin{array}{*{20}c}
              0                  \hfill & 0                    \hfill \\
              - \overset{(1)}{r} \hfill & - \overset{(1)}{R_s} \hfill \\
            \end{array} }} \right) 
            + O (\epsilon^2)
$$ \endgroup
with $J_* = 2\,\parallel \Omega \parallel^2 - div\,g -  A \cdot curl\,\Omega$, hence the scalar curvature expansion:
\begin{equation}
   R = Tr\,(G^{-1} R') = - \epsilon\,(I 
                                      + Tr (\overset{(1)}{R_s})) 
                                      + O (\epsilon^2)\ ,
\label{R = - epsilon (I + Tr (R(1)_s)) + O (epsilon^2)} 
\end{equation}
where:
\begin{equation}
   I = div\,g - 2\,\parallel \Omega \parallel^2 + 2\,A \cdot curl\,\Omega \ ,
\label{I = div g - 2 norm (Omega)^2 + 2 A cdot curl Omega} 
\end{equation}
is a Galilean joint invariant of the components $g, \Omega$ of the Galilean connexion and the potentials $\phi, A$ (\cite{AffineMechBook}, Theorem 16.6). Indeed, taking into account  (\ref{C = (1 -A)}) and (\ref{0 order R'}), it is verified that :
$$ I = - C^T \overset{(0)}{R'} C \ ,
$$
then $I$ is a Galilean invariant because $C$ represents the Galilean vector $\bm{C}$ and $\overset{(0)}{R'}$ represents the Galilean 2-covariant tensor $\overset{(0)}{\bm{R}'}$. 

It is worth to remark the zero order term in (\ref{R = - epsilon (I + Tr (R(1)_s)) + O (epsilon^2)}) is null. We need to develop the last term of order one, owing to (\ref{1st order R'_(lm) = }):
$$ Tr (\overset{(1)}{R_s}) = \overset{(1)\quad}{R'_{ll}} 
     = D_{lj} \Omega^j_m - \dfrac{\partial D_{ll}}{\partial t}
                                 - \dfrac{\partial B_l}{\partial x^l} 
                                 + D_{mj} \Omega^j_l 
                                  + \dfrac{\partial}{\partial x^j} (A^j D_{ll})
                                  - \dfrac{\partial}{\partial x^l} (A^j D_{jl})
$$
The first and fourth terms vanish because they are the trace of the product of a symmetric matrix and a skew-symmetric one. The second and third terms read, taking into account (\ref{B_i = }) and (\ref{D_(ij) = }):
$$ - \dfrac{\partial D_{ll}}{\partial t}
   - \dfrac{\partial B_l}{\partial x^l} 
   = - \dfrac{\partial}{\partial t} (div\,A)
   - div\,(grad\,\phi - \Omega \times A)
$$
or, taking into account:
\begin{equation}
    div\, (u \times v) =   v \cdot curl\,u - u \cdot curl\,v\ ,
\label{div (u times v) = } 
\end{equation}
one has:
$$ - \dfrac{\partial D_{ll}}{\partial t}
   - \dfrac{\partial B_l}{\partial x^l} 
   = div\,(- grad\,\phi - \dfrac{\partial A}{\partial t}) 
     +  A \cdot curl\,\Omega - \Omega \cdot curl\,A\ ,
$$
that gives, owing to (\ref{Galilean gravitation potentials}):
$$ - \dfrac{\partial D_{ll}}{\partial t}
   - \dfrac{\partial B_l}{\partial x^l} 
   = div\,g - 2\,\parallel \Omega \parallel^2 + A \cdot curl\,\Omega\ .
$$
Let us expand the two latter terms:
\begin{eqnarray}
\dfrac{\partial}{\partial x^j} (A^j D_{ll}) - \dfrac{\partial}{\partial x^l} (A^j D_{jl})
 & = & A^j \left( \dfrac{\partial D_{ll}}{\partial x^j} - \dfrac{\partial D_{jl}}{\partial x^l} \right) \nonumber\\
 &   &  + \dfrac{\partial A^j}{\partial x^j}\,D_{ll} - \dfrac{\partial A^j}{\partial x^l}\,D_{jl}\ ,
\label{1st term of 2nd term } 
\end{eqnarray}  
where the two former terms become, owing to (\ref{Galilean gravitation potentials}) and (\ref{D_(ij) = }):
\begin{eqnarray}
     A^j \left( \dfrac{\partial D_{ll}}{\partial x^j}  - \dfrac{\partial D_{jl}}{\partial x^l} \right)
  & = & A^j \left( \dfrac{\partial^2 A_l}{\partial x^j \partial x^l}  
        - \frac{1}{2}\dfrac{\partial^2 A_l}{\partial x^l \partial x^j}  
        - \frac{1}{2}\dfrac{\partial^2 A_j}{\partial x^l \partial x^l} \right) \nonumber\\
  & = & \frac{1}{2} \left( \dfrac{\partial}{\partial x^j} (div\,A) 
        - \triangle A_j \right)\,A^j \nonumber\\
  & = & \frac{1}{2} \left( grad\, (div\,A) - \triangle A \right) \cdot A \nonumber\\
  &  =  & \frac{1}{2} \left( curl\,(curl\,A) \right) \cdot A 
   =  A \cdot curl\,\Omega\ .
\end{eqnarray}
Owing to (\ref{Galilean gravitation potentials}) and (\ref{D_(ij) = }), the two latter terms of (\ref{1st term of 2nd term }) become:
$$ 2\,I_1 = \dfrac{\partial A^j}{\partial x^j}\,D_{ll} - \dfrac{\partial A^j}{\partial x^l}\,D_{jl}
 = (div\,A)^2 - Tr\,\left(\dfrac{\partial A}{\partial x}\,D \right)
$$
$$ 2\,I_1 = (div\,A)^2 - Tr\,\left((D + j (\Omega))\,D \right)
$$ 
The latter term vanishes because it is the trace of the product of a skew-symmetric  matrix and a symmetric one, hence:
$$ 2\,I_1 = (Tr\,D)^2 - Tr\,\left(D^2 \right) = - Tr\,\left[(D - (Tr\,D)\,1_{\mathbb{R}^3})^2\right]\ .
$$
Let us show that this quantity is invariant under Galilean transformations. Indeed, owing to its transformation law (\ref{phi' & A'}), one has:
$$  \frac{\partial A'}{\partial x'} = \frac{\partial}{\partial x}(R^T (A + u))\,R
                                  = R^T \frac{\partial}{\partial x}(A + u)\,R\ ,
$$
Also, differentiating (\ref{eq29}) with respect to $x$ gives:
$$   \frac{\partial u}{\partial x} = j (\omega)\ ,\qquad grad_{\,x}\,u = - j (\omega)\ ,
$$
taking into account the two previous identities, one has::
$$ D' = R^T (D + grad_s u) \,R = R^T D\, R\ ,
$$
leading to:
$$ Tr\,D' = Tr\,D,\qquad Tr\,(D'^2) = Tr\,(D^2)\ ,
$$
that proves $I_1$ is a Galilean invariant. Finally, it holds:
$$ Tr (\overset{(1)}{R_s}) = I + 2\,I_1\ ,
$$
and the scalar curvature expansion (\ref{R = - epsilon (I + Tr (R(1)_s)) + O (epsilon^2)}) becomes:
$$ R =  - 2\,\epsilon\,(I 
                                      + I_1) 
                                      + O (\epsilon^2)\ ,
$$
or in details:
\begingroup\makeatletter\def\f@size{9}\check@mathfonts
\begin{equation}
 R  = - \epsilon\,\left\lbrace 2\,(div\,g - 2\,\parallel \Omega \parallel^2 + 2\,A \cdot curl\,\Omega) 
                                      - Tr\,\left[(D - (Tr\,D)\,1_{\mathbb{R}^3})^2\right]\right\rbrace 
                                      + O (\epsilon^2)\ .
\label{weak field approximation up to the 1st order} 
\end{equation}  \endgroup
It is a Galilean invariant.

\section{Functional due to the geometry}
\label{Section Functional due to the geometry} 

According to (\cite{Souriau 1964}, p. 341, Formula (35.12)), the term of Hilbert-Einstein functional due to the space-time curvature is:
$$ p_G \sqrt{- \det G} = - (p_0 + a\,R)\,\sqrt{- \det G}\ ,
$$
where $p_0$ and $a$ are constants to identify and $R$ is the scalar curvature approximated by (\ref{weak field approximation up to the 1st order}), that leads to:
\begingroup\makeatletter\def\f@size{8}\check@mathfonts
\begin{eqnarray}
p_G \sqrt{- \det G} & = & - c\,\left[   p_0 - 2 \, a\,\epsilon \,
\left\lbrace 2\,(div\,g - 2\,\parallel \Omega \parallel^2 + 2\,A \cdot curl\,\Omega) 
                                         - Tr\,\left[(D - (Tr\,D)\,1_{\mathbb{R}^3})^2\right]\right
\rbrace
                                            \right]  \, \nonumber\\
                              &    &      \left(1 + \epsilon\,\left(\phi 
                                                       + \frac{1}{2}\,\parallel A \parallel^2
                                            \right)
                            \right)\ ,
\label{p_G sqrt( - det G) = 1st order approximation} 
\end{eqnarray} \endgroup
or, in short:
$$   p_G \sqrt{- \det G}  =  - c\, [   p_0 - 2 \, a\,\epsilon \,  \left\lbrace I + I_1\right\rbrace ]  \, (1 + \epsilon \, I_0 )\ .
$$
Every term of the expansion of this expression is a Galilean invariant.

The variation of Hilbert-Einstein functional (\ref{Hilbert-Einstein functional}):
\begingroup\makeatletter\def\f@size{9}\check@mathfonts
\begin{equation}
c\, \int \,\left\lbrace \left[ - p_0 + 2 \, a\,\epsilon \,( I + I_1) \right]\,(1 + \epsilon\,I_0) 
+ \rho\,\epsilon^{-1} 
                - \rho\,\left(\dfrac{1}{2}\,\parallel v \parallel^2 + A \cdot v - \phi\right) 
      \right\rbrace\,d^4 X
\label{Hilbert-Einstein functional 2} 
\end{equation} \endgroup
with respect to the Galilean potential $\phi$ gives:
$$ I + I_1                               = - \frac{\rho}{2\,a\,\epsilon^2} + \frac{p_0}{2\,a\,\epsilon} \ ,
$$
or explicitly:
$$ div\,g - 2\,\parallel \Omega \parallel^2 + 2\,A \cdot curl\,\Omega
                                      - \frac{1}{2}\,Tr\,\left[(D - (Tr\,D)\,1_{\mathbb{R}^3})^2\right]
                                      = - \frac{\rho}{2\,a\,\epsilon^2} + \frac{p_0}{2\,a\,\epsilon} \ .
$$
To identify the unknown constants, we consider the particular case without spinning ($A = \Omega =  0$ hence $D = 0$ by (\ref{D_(ij) = })):
$$ div\,g = - \frac{\rho}{2\,a\,\epsilon^2} + \frac{p_0}{2\,a\,\epsilon} \ .
$$
By comparison to:
$$ div\,g = - 4\,\pi\,k_N \rho + \Lambda \ ,
$$
that generalizes Poisson's equation where $k_N$ is the gravitational constant and the extra term is due to the cosmological constant $\Lambda$,
we put:
$$ \frac{1}{a} = 8\,\pi\,k_N \epsilon^2 = \frac{8\,\pi\,k_N}{c^4},\qquad 
     p_0 = 2\,a\,\Lambda\,\epsilon = \frac{2\,a\,\Lambda}{c^2}\ .
$$
in (\ref{Hilbert-Einstein functional 2}):
\begingroup\makeatletter\def\f@size{9}\check@mathfonts
$$ c\, \int \,\left\lbrace  \frac{\epsilon^{-1}}{4\,\pi\,k_N}  \left[  I + I_1 - \Lambda) \right]\,(1 + \epsilon\,I_0) 
+ \rho\,\epsilon^{-1} 
                - \rho\,\left(\dfrac{1}{2}\,\parallel v \parallel^2 + A \cdot v - \phi\right) 
      \right\rbrace\,d^4 X
$$ \endgroup
The variation of the new functional with respect to the Galilean potentials $\phi$ and $A$ gives the Galilean gravitation field equations:
\begin{equation}
\begin{tabular}{|l l|}
\hline 
& \\
    $ I + I_1  - \Lambda                                                                   = - 4\,\pi\,k_N \rho,   $ & \\
& \\
\hline 
\end{tabular}
\label{Einstein equation for the Galilean gravitation time} 
\end{equation}
\begin{equation}
\begin{tabular}{|l l|}
\hline 
& \\
    $ - 2\,(\epsilon^{-1} + I_0)\,curl\,\Omega - (I + I_1 - \Lambda)\,A = - 4\,\pi\,k_N \rho\,v\ .   $ & \\
& \\
\hline 
\end{tabular}
\label{Einstein equation for the Galilean gravitation space} 
\end{equation}
It is clear that the right hand side members are the components of the $4$-flux $- 4\,\pi\,k_N \rho\,U$. Because $x_0$ is a Galilean invariant, the spacetime gradient is modified under a Galilean transformation $P$ according to:
$$ \frac{\partial x_0}{\partial X'} = \frac{\partial x_0}{\partial X} \,\frac{\partial X}{\partial X'} 
                                                   =  \frac{\partial x_0}{\partial X} \,P
$$
Taking into account (\ref{Galilean transfo}), (\ref{partial s' / partial X = (- F^(-1) v, F^(-1))}) and (\ref{v'=R^T (v - u)}), the transformation law of the deformation gradient is $ F' = R^T F$ then $\det F$ is a Galilean invariant. As the $4$-velocity $U$ is a Galilean vector, the $4$-flux $- 4\,\pi\,k_N \rho\,U$ is a Galilean vector too. On the other hand, let us verify that the $4$-column:
$$ V  =\left( \begin{array} {c}
                   I + I_1  - \Lambda  \\
                   - 2\,(\epsilon^{-1} + I_0)\,curl\,\Omega - (I + I_1 - \Lambda)\,A\\
       \end {array} \right)\ ,
$$
represents a Galilean vector. Indeed, it holds:
$$ d \Omega' = \frac{\partial \Omega'}{\partial X}\,dX = \frac{\partial \Omega'}{\partial X}\,P\,dX'\ .
$$ 
On the other hand, $v'$ being seen as a function of $X'$ through the coordinate change $X \longmapsto X'$, one has:
$$ d \Omega' = \frac{\partial \Omega'}{\partial X'}\,dX' \ .
$$ 
$dX'$ being arbitrary, we obtain by comparing the previous relations:
$$ \frac{\partial \Omega'}{\partial X'} = \frac{\partial \Omega'}{\partial X}\,P\ .
$$
Taking into account (\ref{Galilean transfo}), one has:
$$\frac{\partial \Omega'}{\partial t'} = \frac{\partial \Omega'}{\partial t} 
    + \frac{\partial \Omega'}{\partial x}\,u,\qquad
     \frac{\partial \Omega'}{\partial x'} = \frac{\partial \Omega'}{\partial x}\,R\ ,
$$
and, owing to (\ref{transformation law Omega}):
$$ \frac{\partial \Omega'}{\partial x'} = R^T \frac{\partial \Omega}{\partial x}\,R\ .
$$
Using (\ref{j(R^T u)}) and (\ref{defi curl})  leads to the transformation law of of the curl of $\Omega$:
$$ curl_{\,x'}\,\Omega' = R^T curl_{\,x}\,\Omega\ ,
$$
and, owing to invariance of $(I + I_1)$ and the transformation law (\ref{phi' & A'}) of $A$, the $4$-column in another Galilean coordinate system:
$$ V'  =\left( \begin{array} {c}
                   I' + I'_1  - \Lambda  \\
                   - 2\,(\epsilon^{-1} + I'_0)\,curl_{\,x'}\,\Omega' - (I' + I'_1 - \Lambda)\,A'\\
       \end {array} \right) \ ,
$$
is given by:
$$ V'  =\left( \begin{array} {c}
                   I + I_1  - \Lambda  \\
                   - 2\,(\epsilon^{-1} + I_0)\,R^T curl_{\,x}\,\Omega - (I + I_1 - \Lambda)\,R^T (A + u)\\
       \end {array} \right)\ ,
$$
that is the transformation law $V' = P^{-1} V$ of Galilean vectors with the Galilean transformation (\ref{Galilean transfo}). Hence we verified the consistency of the Galilean gravitation field equations (\ref{Einstein equation for the Galilean gravitation time}) and (\ref{Einstein equation for the Galilean gravitation space}) in the sence that they mean the equality of two Galilean vectors.

\section{Asymptotic expansion of the solution}
\label{Section Asymptotic expansion of the solution} 

Considering in this Section that the motion of the matter is known, then the fields $\rho$ and $v$ are given, equations (\ref{Einstein equation for the Galilean gravitation time}) and (\ref{Einstein equation for the Galilean gravitation space}) can be used to determine the Galilean gravitation due the presence of the matter. As the equations are non linear, the solutions is expanded into:
$$ \phi =        \overset{(0)}{\phi}
                             + \epsilon\,\overset{(1)}{\phi}
                             + O (\epsilon^2),\qquad
      A =        \overset{(0)}{A}
                             + \epsilon\,\overset{(1)}{A}
                             + O (\epsilon^2)\ .
$$
The corresponding gravitation components are expanded as:
$$ g =        \overset{(0)}{g}
                             + \epsilon\,\overset{(1)}{g}
                             + O (\epsilon^2),\qquad
      \Omega =        \overset{(0)}{\Omega}
                             + \epsilon\,\overset{(1)}{\Omega}
                             + O (\epsilon^2)\ ,
$$
with:
\begin{equation}
     \overset{(i)}{g} = - grad\,\overset{(i)}{\phi} - \dfrac{\partial \overset{(i)}{A}}{\partial t},\qquad  
     \overset{(i)}{\Omega} = \frac{1}{2}\,curl\,\overset{(i)}{A}\ ,
\label{Galilean gravitation potentials at order i} 
\end{equation}
verifying:
\begin{equation}
        curl\,\overset{(i)}{g} + 2 \,\frac{\partial \overset{(i)}{\Omega}}{\partial t} = 0,\qquad 
        div\,\overset{(i)}{\Omega} = 0\ ,
\label{curl g + 2 (partial Omega / partial t) = 0 & div Omega = 0 at order i} 
\end{equation}
because of (\ref{curl g + 2 (partial Omega / partial t) = 0 & div Omega = 0}).  

\subsection{Potential vector at order zero}

Equation (\ref{Einstein equation for the Galilean gravitation space}) gives at order $\epsilon^{-1}$:
\begin{equation}
    curl\,\overset{(0)}{\Omega} = 0\ .
\label{curl A (0) = 0} 
\end{equation}
Besides, owing to (\ref{curl g + 2 (partial Omega / partial t) = 0 & div Omega = 0 at order i}):
$$ div\,\overset{(0)}{\Omega} = 0\ ,
$$
and the field is harmonic:
$$   \triangle \overset{(0)}{\Omega} 
       = grad\,(div\,\overset{(0)}{\Omega}) - curl\,(curl\,\overset{(0)}{\Omega}) = 0\ .
$$
Assuming that $\overset{(0)}{\Omega}$ is bounded, it is uniform because of the maximum principle:
\begin{equation}
    \overset{(0)}{\Omega} = - \omega (t)\ ,
\label{A (0) = - omega (t)} 
\end{equation}
where $\omega$ is an arbitrary smooth function of the time. The sign is purely conventional.  Although the point of  view is rather different from the GGT, it is worth to observe that the previous condition is similar to the one deduced from Ehlers conditions \cite{Andringa Bergshoeff 2011}. Besides, integrating (\ref{Galilean gravitation potentials at order i}), the general solution is:
$$ \overset{(0)}{A} = - \omega (t) \times x + u_0 (t) \ ,
$$
where $u_0$ is an arbitrary smooth function. Considering a solution $x_0$  of the linear ODE:
$$ \dot{x}_0 - \omega \times x_0 = u_0
$$ 
We recover minus the velocity of transport (\ref{eq29}):
\begin{equation}
   \overset{(0)}{A} = - ( \omega \times (x - x_0)+ \dot{x}_0) = - u\ ,
\label{A (0) = - u} 
\end{equation}
hence the gradient of $\overset{(0)}{A}$ is skew-symmetric 
$$ grad\, \overset{(0)}{A} = - grad\, u = j (\omega)
$$ 
and by (\ref{D_(ij) = }), one has:
$$ \overset{(0)}{D} = 0,\qquad \overset{(0)}{I}_1 = 0\ .
$$
 Using (\ref{div (u times v) = }), remark also that $\overset{(0)}{A}$ is free divergence:
$$ div \, \overset{(0)}{A} = 0
$$

\subsection{Scalar vector at order zero}

Owing to (\ref{Galilean gravitation potentials at order i}), we have:
$$ div\,\overset{(0)}{g}  = - \triangle \overset{(0)}{\phi} \ .
$$
Then taking into account (\ref{I = div g - 2 norm (Omega)^2 + 2 A cdot curl Omega}) and (\ref{curl A (0) = 0}), one has:
$$ \overset{(0)}{I}  = div\,\overset{(0)}{g} 
                                  - 2\,\parallel \overset{(0)}{\Omega} \parallel^2 
                                  + 2\,\overset{(0)}{A} \cdot curl\,\overset{(0)}{\Omega }  
                              = - \triangle \overset{(0)}{\phi}  
                                  - 2\,\parallel \overset{(0)}{\Omega} \parallel^2  
$$
The equation (\ref{Einstein equation for the Galilean gravitation time}) at order $0$ is reduced to:
$$ \triangle\,\overset{(0)}{\phi} 
                                     = 4\,\pi\,k_N \rho - 2\,\parallel \overset{(0)}{\Omega} \parallel^2  - \Lambda\ .
$$
As it is linear, its general solution is the sum:
$$ \overset{(0)}{\phi} = \phi_N + \phi_\Omega + \phi_\Lambda
$$
of the contribution $\phi_N$ due to the matter, $\phi_\Omega$ due to the spinning and $\phi_\Lambda$ due to the cosmological constant:
$$  \triangle\,\phi_N
                                     = 4\,\pi\,k_N \rho,\qquad  
  \triangle\,\phi_\Omega
                                     = - 2\,\parallel \overset{(0)}{\Omega} \parallel^2 , \qquad
 \triangle\,\phi_\Lambda 
                                     =   - \Lambda\ .
$$
The field $\phi_N$, solution of Poisson's equation is well known. The field $\phi_\Omega$ is also within reach. Comparing (\ref{phi = - (1/2) norm(u)^2 & A = - u}) and (\ref{A (0) = - u}), we try the test function:
$$ \phi_\Omega = - \frac{1}{2}\,\parallel u \parallel^2 
$$
of which the gradient is:
$$ grad\,\phi_\Omega = - (grad\,u)\,u = j (\omega)\,u 
                                    = - \overset{(0)}{\Omega} \times u
$$
then, taking into account (\ref{div (u times v) = }) and (\ref{A (0) = - u}), one has:
$$ \triangle \phi_\Omega  = div\,(grad\,\phi_\Omega) 
                                          = - \overset{(0)}{\Omega} \cdot curl \overset{(0)}{A} 
                                          = - 2\, \parallel \overset{(0)}{\Omega} \parallel^2
$$
Finally, we check that:
$$ \phi_\Lambda = - \frac{1}{6}\,\Lambda\,\parallel x \parallel^2 
$$ 
In a nutshell, the complete solution is:
$$ \overset{(0)}{\phi} (x,t) = - \int \frac{k_N \,\rho (\bar{x},t)}{\parallel x - \bar{x} \parallel}\,d\mathcal{V} (\bar{x})
                                             - \frac{1}{2}\,\parallel u (x,t) \parallel^2 
                                             - \frac{1}{6}\,\Lambda\,\parallel x \parallel^2  \ .
$$

\subsection{Inertial frames}

For the following discussion, we remove provisionally the effect of the cosmological constant.  The result of the previous calculus is the expression of the potentials of the Galilean gravitation at the classical approximation
\begin{equation}
\phi = \overset{(0)}{\phi} = - \int \frac{k_N \,\rho (\bar{x},t)}{\parallel x - \bar{x} \parallel}\,d\mathcal{V} (\bar{x})
                                             - \frac{1}{2}\,\parallel u (x,t) \parallel^2  , \qquad 
     A = \overset{(0)}{A} =  - u (x,t)\ .
\label{potentials in inertial frames} 
\end{equation}
where $u$ is the velocity of transport (\ref{eq29}). We call inertial frames the reference frames in which Newton's law of universal gravitation is true: 
$$ g            = - \int \frac{k_N \,\rho (\bar{x},t)}{\parallel x - \bar{x} \parallel^2}\,
                         \frac{x - \bar{x}}{\parallel x - \bar{x} \parallel}\,
                         d\mathcal{V} (\bar{x}),\qquad
    \Omega = 0
$$
and, in absence of gravitation, the particles are in uniform straight motion. For this reason, we proposed in \cite{AffineMechBook} to call them Newtonian frames instead of inertial frames. As remarked before, the inertial forces are generated by the potentials (\ref{phi = - (1/2) norm(u)^2 & A = - u}). Then the potentials are given in inertial frames $X$ by 
(\ref{potentials in inertial frames}) with $u = 0$. 

Consequently, Poisson's equation is valid only in inertial frames and, because $\Omega = 0$, the Ricci tensor (\ref{0 order R'}) is reduced to:
$$   R'  = \left( {{\begin{array}{*{20}c}
              - div\,g \hfill & 0^T \hfill \\
              0                            \hfill & 0                \hfill \\
            \end{array} }} \right)
           = \left( {{\begin{array}{*{20}c}
              4\,\pi\,k_N \rho \hfill & 0^T \hfill \\
              0                            \hfill & 0                \hfill \\
            \end{array} }} \right)\ .
$$
We recover Trautman's form (\ref{R' = 4 pi k_N rho tau otimes tau}) of Ricci tensor but which is not Galilean covariant because $div\, g$ is not invariant.

Conversely, inverting (\ref{transformation law of g - 2 Omega x v}), the gravitation in a non inertial frame $X'$ is given by:
$$   m\,(g' - 2\,\Omega' \times v') = R^T (m\,g - m\,a_t)
$$
where the former term of the right hand member represents the effect ot the universal attraction and the latter one represents the change of reference frame, through the inertial forces (\ref{inertial forces}). According to Einstein equivalence principle, the laws of the Nature are such that it is impossible to distinguish between the effect of the universal attraction and the one of the frame change. 

As example, let us consider a rapidly rotating star of centre  $ x_0 = 0$  with constant rotation rate $\Omega$. Taking into account (\ref{eq29}) and (\ref{A (0) = - omega (t)}), (\ref{potentials in inertial frames}) gives:
$$\phi = - \int \frac{k_N \,\rho (\bar{x},t)}{\parallel x - \bar{x} \parallel}\,d\mathcal{V} (\bar{x})
                                             - \frac{1}{2}\,\parallel \Omega \times x \parallel^2  , \qquad 
     A = \Omega \times x \ .
$$
recovering the expression of the scalar potential used in \cite{Lignieres Rieutord 2006}. By (\ref{Galilean gravitation potentials}),  the corresponding gravity component is:
$$ g            = - \int \frac{k_N \,\rho (\bar{x},t)}{\parallel x - \bar{x} \parallel^2}\,
                         \frac{x - \bar{x}}{\parallel x - \bar{x} \parallel}\,
                         d\mathcal{V} (\bar{x})
                       - \Omega \times (\Omega \times x)
$$
where the last term is the centrifugal acceleration.

\subsection{Potential vector at order one}

Resuming  the asymptotic expansion where we left off, Equation (\ref{Einstein equation for the Galilean gravitation space}) gives at order $0$:
$$ 2\,curl\,\overset{(1)}{\Omega} =  4\,\pi\,k_N \rho\,v\ ,
$$
Owing to (\ref{Galilean gravitation potentials at order i}):
$$ 2\,curl\,\overset{(1)}{\Omega} =  curl\,(curl\,\overset{(1)}{A})
  = grad\,(div\,\overset{(1)}{A}) - \triangle \overset{(1)}{A} \ .
$$
By a gauge transformation, we can choose the potential divergence free, hence the equation:
$$ \triangle \overset{(1)}{A} = - 4\,\pi\,k_N \rho\,v\ ,
$$
of which the solution is:
$$ \overset{(1)}{A} (x,t) = \int \frac{k_N \,\rho (\bar{x},t)\,v (\bar{x},t)}{\parallel x - \bar{x} \parallel}\,d\mathcal{V} (\bar{x})\ .
$$
We recover the same dominant term of the PPN expansion as in \cite{Peter Uzan 2012}. In short, the general solution is:
\begingroup\makeatletter\def\f@size{10}\check@mathfonts
$$ \phi (x,t) = - \int \frac{k_N \,\rho (\bar{x},t)}{\parallel x - \bar{x} \parallel}\,d\mathcal{V} (\bar{x})
                                             - \frac{1}{2}\,\parallel u (x,t) \parallel^2 
                                             - \frac{1}{6}\,\Lambda\,\parallel x \parallel^2 
                                             + O \left(\frac{1}{c^2}\right)\ ,
$$ 
$$ A (x,t) =  - u (x,t) + \frac{1}{c^2}\int \frac{k_N \,\rho (\bar{x},t)\,v (\bar{x},t)}{\parallel x - \bar{x} \parallel}\,d\mathcal{V} (\bar{x}) + O \left(\frac{1}{c^4}\right)\ .
$$ \endgroup

\section{Conclusions}

In this work, we revisited the old problem of derivating the gravitation field equations for the Galilean relativity. The novelty of our point of view lies in considering $c$ expansion of the fields as in PPN approach but taking due care that every term of the expansion has the Galilean covariance. Not only we obtain a condition for the scalar potential but also the three missing equations allowing to determine the vector potential. These four equations has the expected Galilean covariance. These equations being non linear, we obtained solutions by asymptotic expansion. The analyse of the solution reveals that Poisson's equation is valid only in the inertial frames. Then we generalized it for arbitrary Galilean frames, inertial and non inertial. Besides, we showed that the second component of the Galilean gravitation called spinning is, at least with an error of the order of $c^{-2}$,  irrotational, a simple result but not easy to deduce. As the field is divergence free, it is harmonic and, under a reasonnable hypothesis of boundedness, it is uniform.

As regards the prospects, some extensions are naturally open to us. Firstly, although the calculus are already cumbersome at the first order, the final expression of the functional and the field equations are rather compact. It seems possible to tackle the order two to obtain finer corrections. Secondly, another topics of interest is to enrich the metrics while conserving the Galilean feature of the terms of the expansion. The simplest enrichment consists to add another degree of freedom $\psi$ to the gravitation by considering the covariant metrics:
$$    G =\left( {{\begin{array}{cc}
               c^2 + 2\,\phi \hfill &   -A^T             \hfill \\
               - A           \hfill & - \psi\,1_{\mathbb{R}^3} \hfill \\
   \end{array} }} \right)\ ,
$$
According to the transformation law (\ref{transformation law of a & w & M}), the extra potential $\psi$ must be a Galilean invariant. However, this kind of approach is relevant only to compare to cosmological data at large enough scales. In contrast, to approximate correctly Schwarzschild solution  we need a more general form:
$$    G =\left( {{\begin{array}{cc}
               c^2 + 2\,\phi \hfill &   -A^T             \hfill \\
               - A           \hfill & - M \hfill \\
   \end{array} }} \right)\ ,
$$
with a symmetric but anisotropic matrix $M$, that paves the way to a large class of  intermediate models with more than $5$ potentials towards the GR. Thirdly, it would be interesting to search exact solutions of the non linear gravitation field equations, at least with additional symmetries, for instance spherical. Finally, we thinks this kind of approach could be useful for the Cosmology, for instance modelling the distribution of galaxies at local scale as done in \cite{Fliche Triay 2006}.

\section{Aknowledgement} 

I would like to acknowledge the Mainz Institute for Theoretical Physics (MITP) for enabling me to improve significantly this work during the workshop \textquote{Applied Newton-Cartan Geometry} (APPNC 2018).

\end{document}